\documentclass[notitlepage,superscriptaddress, prl, aps, floatfix, twocolumn]{revtex4}
\usepackage{graphicx,bm}
\usepackage{subfigure}
\usepackage{comment}
\usepackage{amssymb}
\usepackage{amsmath}
\usepackage{verbatim}
\usepackage{my_symbols}
\usepackage{amsfonts}
\usepackage{graphicx}
\usepackage{xcolor}
\usepackage{CJK}
\usepackage[normalem]{ulem}%
\providecommand{\U}[1]{\protect\rule{.1in}{.1in}}
\providecommand{\U}[1]{\protect\rule{.1in}{.1in}}
\RequirePackage[T1]{fontenc}
\RequirePackage{times}

\newcommand\rmv{\bgroup\markoverwith {\textcolor{magenta}{\rule[0.5ex]{2pt}{0.6pt}}}\ULon}

\begin{document}
\begin{CJK*}{UTF8}{gbsn} 
\title{Transport between metals and magnetic insulators}
\date{\today}
\author{Jiang Xiao (萧江)}
\affiliation{Department of Physics and State Key Laboratory of Surface Physics, Fudan
University, Shanghai 200433, China}
\affiliation{Collaborative Innovation Center of Advanced Microstructures, Fudan University,
Shanghai, 200433, China}
\author{Gerrit E. W. Bauer}
\affiliation{Institute for Materials Research and WPI-AIMR, Tohoku University, Sendai, Japan}
\affiliation{Kavli Institute of NanoScience, Delft University of Technology, Delft, The Netherlands}
\begin{abstract}
We derive the Onsager response matrix of fluctuation-mediated spin-collinear
transport through a ferromagnetic insulator and normal metal interface driven
by a temperature difference, spin accumulation, or magnetic field. We predict
magnon-squeezing spin currents, magnetic field-induced cooling (magnon Peltier
effect), temperature induced magnetization (thermal magnetic field) as well as
universal spin Seebeck/Peltier coefficients.
\end{abstract}
\maketitle
\end{CJK*}

Finite temperature effects on the transport properties of magnetic nanostructures \cite{pulizzi_spintronics_2012} attracts considerable attention since the discovery of the spin Seebeck effect \cite{uchida_observation_2008, uchida_spin_2010, jaworski_observation_2010} that thwarts conventional thermoelectrics. Of special interest are heterostructures of magnetic insulators such as yttrium iron garnets (YIG) with heavy normal metals such as Pt, where the latter, via the inverse spin Hall effect, function as spin current detectors. Here we report a linear response approach to thermal transport through interfaces between ferro- or ferrimagnetic insulators (FI) and normal metals (N) that extends our treatment of the spin Seebeck effect \cite{xiao_theory_2010} to the spin Peltier effect and leads to the prediction of, e.g., a magnon Peltier effect and its Onsager reciprocal, a thermal effective field.

The Landau-Lifshitz-Gilbert equation for the dynamics of a magnetization in an effective magnetic field $\bB_{0}$
\begin{equation}
    \dbm = -\gamma\mb\times\smlb{\bB_0 + \bb} +\alpha \mb\times\dbm 
    \label{eqn:llg}
\end{equation}
is based on the assumption that the modulus of the spatiotemporal magnetization texture $\bM(\br,t)$ is constant, \ie $\bM(\br,t) = M_{s}\mb(\br,t) $ and $\abs{\mb}=1$, which is valid at temperatures sufficiently below that of the magnetic phase transition. The LLG predicts a temperature-induced reduction of the time-averaged equilibrium magnetization by considering a stochastic magnetic field $\bb(\br,t)$ that induces thermal fluctuations of $\mb$ around the equilibrium direction. Thermal noise is characterized by the spatiotemporal correlation function $\avg{ b_i(\br,t)b_j(\br',t')} $ that by the Fluctuation-Dissipation Theorem (FDT) can be expressed in terms of an integral over the Bose-Einstein distribution function of the magnon excitations and proportional to the Gilbert damping constant $\alpha$ \cite{brown_thermal_1963}. 
Microscopically, the magnetization noise in insulating ferromagnets is caused by the magnetoelastic interaction that couples and equilibrates the magnetic and elastic order parameters. At interfaces to metals, spin pumping induces an additional energy and angular momentum dissipation that increases the effective damping and the magnetic fluctuations \cite{foros_magnetization_2005,foros_resistance_2007,xiao_charge_2009}.

Spin accumulations in the normal metal at interfaces to ferromagnets with transverse spin polarization generate spin-transfer torques \cite{slonczewski_current-driven_1996,berger_emission_1996}, while the longitudinal ones have at zero temperature no effect on the magnetization. One might therefore, naively, expect that the magnetization of the insulator in FI$|$N bilayers is inert without outside spin injection or non-collinear magnetic fields. However, spin collinear transport phenomena in FI$|$N systems exist at finite temperatures by the magnetic thermal fluctuations that allow a longitudinal spin accumulation in N to act on instantaneous transverse magnetization components. The ensemble/time average of the thus induced spin currents is polarized along the equilibrium magnetization direction.

Perturbation on a system at thermal equilibrium generates a response or ``current'' that is proportional to the ``force'' when the latter is sufficiently weak (Ohm's Law). In the presence of multiple forces and currents cross-correlations exist, thermoelectrics being a prominent example. The linear response of such a system is then described by a ``conductance'' matrix that relates forces and currents, which possesses a fundamental symmetry referred to as Onsager reciprocity \cite{onsager_reciprocal_1931} that is very useful in spintronics \cite{johnson_thermodynamic_1987}.   
Here we establish the Onsager matrix for transport through a normal metal and a ferromagnetic insulator contact (N$|$FI) that is actuated by a spin accumulation in N, temperature difference over the interface, and (pulsed) external magnetic field. Each matrix element represents a different physical phenomenon, of which the spin Seebeck effect is just one \cite{xiao_theory_2010,adachi_linear-response_2011,hoffman_landau-lifshitz_2013}. The extended Onsager matrix discussed in the following has already been implicitly used (without details and with reference to the present work) in the analysis of the spin Peltier effect \cite{flipse_observation_2014} and in modelling spin Seebeck generators \cite{cahaya_spin_2014}. The N$|$FI system has recently been analyzed by Bender \etal \cite{bender_interfacial_2015} using a ``Golden Rule'' treatment of the interface exchange interaction including angle-dependent spin transfer torques and quantum effects. However, this study does not take into account the magnetic field component parallel to the magnetization that is central to the present work. We focus on symmetry conserving perturbations, thereby disregarding deterministic transverse spin accumulations and spin-transfer torques, which is allowed as long as the systems is far from the threshold of charge current-induced magnetic self-oscillations or magnetization reversal. All elements of the response matrix are then scalars. Recently, Nakata \etal \cite{nakata_wiedemann-franz_2015} derived the Onsager matrix for a bilayer of magnetic insulators actuated by inhomogeneous magnetic fields and temperature differences.

\begin{figure}[t] 
    \includegraphics[height=3cm]{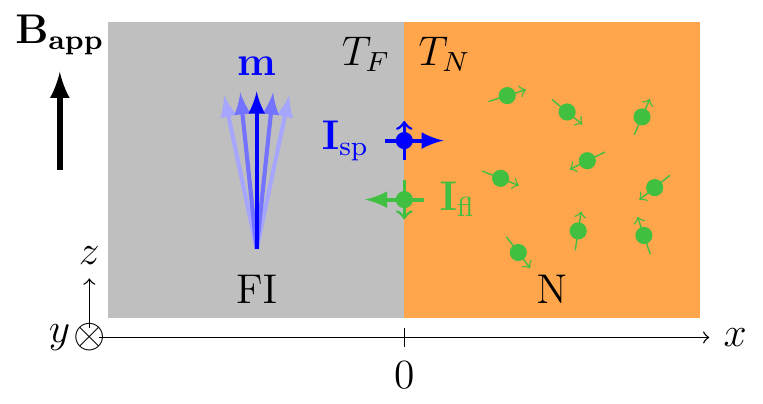} 
    \caption{(Color online) Spin and energy current driven by the thermal bias ($\Delta T=T_{F}-T_{N}$), spin chemical potential $\bV_{s}$, and an external magnetic field $\bB_{\rm app}$ at an FI$|$N interface. } 
    \label{fig:FIN}
\end{figure}

\Figure{fig:FIN} sketchs the ferromagnetic insulator (FI) with normal metal (N) contact. The equilibrium FI magnetization is parallel to the uniaxial anisotropy field $\bB_0 = B_0\hbz =(\omega_{0}/\gamma)\hbz \| \avg{\mb}$. We adopt the ``three reservoir model'' by assuming that thermalization of spin waves in FI and electrons in N is sufficiently efficient. The steady state in the presence of a temperature gradient is then characterized by the temperatures of phonon (FI and N), magnon (FI), and electron (N) systems, thereby disregarding spin dependent electron temperatures in N \cite{dejene_spin_2013}. Transport is generated by the differences in the thermodynamic variables on both sides of the interface. This situation is amenable to magnetoelectronic scattering theory of transport parameterized by the interface scattering matrix. This picture has a wider applicability in circuit theories, in which the interfaces generate boundary conditions between ``bulk regions'' that may be described by quasi-equilibrium distribution functions \cite{brataas_non-collinear_2006}.  For comparison with experiments, the present analysis is a (crucial) building block in simulating entire devices \cite{schreier_magnon_2013,flipse_observation_2014}. 

For simplicity, we consider here the limit of small phonon (Kapitza) heat conductances, which allows us to discard the phonons altogether \cite{cahaya_spin_2014}. The thermodynamic state of the insulating ferromagnet is then characterized by the temperature $T_{F}$ only. The normal metal is at electron and phonon temperature $T_{N}$. We include the option of having long-lived spin accumulation, i.e., a chemical potential difference between the spin up and spin down electrons in the frame of the FI equilibrium magnetization with quantization axis along $\hbz$. The thermodynamic forces are then the external magnetic field $\bB_{\rm app} = B_{\rm app}\hbz = (\omega_{\rm app}/\gamma)\hbz$, longitudinal spin accumulation $\mathbf{V}_{s}=V_{s}\mathbf{\hat{z}}=(\hbar\omega_{s} /2e)\mathbf{\hat{z}}$ (in units of volt) in N, and temperature difference $\Delta T=T_{F}-T_{N}$ across the interface. With $T=(T_{F}+T_{N})/2$, the rate of change of the free energy reads
\begin{equation} 
    \dF = \dS V_{s}+\dQ\frac{\Delta T}{T}-\dM_{z}B_{\rm app},
\label{eqn:F}
\end{equation}
where $\bS(t)=\int_{N}\bs(\br,t)dV=S\hbz$ is the total spin (in units of electric charge $\abs{e}$) in N with spin density $\bs(\br,t)$, $\dQ$ is the heat current entering N, $M_{z}(t)=M_{s}\int_{\rm FI}m_{z}(\br,t)dV$ is the total magnetization in FI with local magnetization component $M_{s}m_{z}(\br,t)$. The volume of the FI is $V=Ad$ with interface area $A$ and thickness $d$. $\dF$ includes contributions from the electrons (first two terms) and the magnetization (last term). Since Onsager symmetry holds when the entropy generation rate $\dot{\cS} = \dF/T$ equals the sum of the product of currents and forces, we identify in \Eq{eqn:F} $V_{s},-\Delta T/T,-B_{\rm app}$ as thermodynamic driving forces with spin current density $j_{s}=(\hbar /2e)\dS/A$, energy current density $j_{Q}=\dQ/A$, and magnetization dynamics $\dM_{z}/\gamma A$, as the conjugated flows, respectively. The linear response coefficient matrix in
\begin{equation} 
    \smatrix{{\dM_z\ov\gamma A} \\ j_s \\ j_Q} = \smatrix{
    L_{mm} & L_{ms} & L_{mQ} \\ 
    L_{sm} & L_{ss} & L_{sQ} \\ 
    L_{Qm} & L_{Qs} & L_{QQ}} 
    \smatrix{\omega_{\rm app} \\ \omega_s \\ {\Delta T\ov T}}
\label{eqn:JV}
\end{equation}
must then satisfy the Onsager reciprocal relations \cite{onsager_reciprocal_1931}, which demand that our response matrix is symmetric. The matrix element $L_{sQ}$ parametrizes the spin Seebeck effect \cite{xiao_theory_2010, adachi_linear-response_2011, hoffman_landau-lifshitz_2013}. Other elements represent various physical effects; $L_{ss}$ is the longitudinal spin conductance, $L_{Qs}$ governs the spin Peltier effect , $L_{QQ}$ is the heat conductance, while the signficance of the new matrix elements $L_{Qm}$ and $L_{mQ}$ is discussed below.

\textit{Macrospin model - } We first treat the macrospin limit in which $\mb = \bM/M$ with $M=M_{s}V$ constant in space and described by the Landau-Lifshitz-Gilbert equation including interface torques:
\begin{equation}
    \dbm = -\gamma\mb\times \bB_{\rm eff} + \alpha \mb \times \dbm
    - \alpha' \omega_s \mb \times (\mb \times \hbz).
\label{eqn:llg0}
\end{equation}
where $\bB_{\rm eff} = \bB_0 + \bB_{\rm app} + \bb$ includes the internal field $\bB_0\|\hbz$ and the applied field $\bB_{\rm app}\|\hbz$. The last term is the instantaneous spin-transfer torque acting on the magnetization by the longitudinal spin accumulation in N \cite{slonczewski_current-driven_1996,berger_emission_1996}. The Gilbert damping parameter $\alpha$ is the sum of the intrinsic $\alpha_{0}$ and $\alpha' = \gamma\hbar g_r/(4\pi M_sd)$, the interface damping \cite{tserkovnyak_enhanced_2002}, where $g_{r}$ is the real part of the spin mixing conductance per unit area of the FI$|$N interface \cite{brataas_non-collinear_2006}. Thermal effects are modeled by the stochastic fields $\bb = \bb_0 + \bb'$, which by the FDT, is associated with the damping $\alpha=\alpha_{0}+\alpha'$ \cite{brown_thermal_1963,foros_noise_2009}. Sufficiently far from the Curie temperature, only the transverse $x$ and $y$ components of $\mb, \bb$, and $\bb'$ matter. To lowest order, the Fourier components of $m_{x,y}$ satisfy $\td{m}_{i}^{\omega}=\sum_{j}\chi _{ij}^{\omega}\gamma\td{b}_{j}^{\omega}$ ($i,j=x,y$), introducing the magnetic susceptibility
\begin{equation}
    \chi^{\omega} = \frac{-(1+\alpha^2)^{-1}}{(\omega-\omega_{0}^{+})(\omega-\omega_{0}^{-})}
    \smatrix{\omega_0-i\alpha\omega & -i\omega \\ i\omega & \omega_0-i\alpha\omega}
\label{eqn:chi}
\end{equation}
with $\omega_{0}^{\pm}=\pm\omega_{0}/(1\pm i\alpha)$. According to the FDT \cite{landau_statistical_1980}, the magnetic fluctuations for $k_{B}T\gg \hbar\omega_{0}$ satisfy
\begin{equation}
    {\avg{\td{m}_i^{\omega}\td{m}_j^{\omega'}}} = 
    \frac{\gamma\hbar k_BT_F}{M} \frac{\smlb{\chi - \chi^\dagger}_{ij}}{i\hbar\omega}
    \delta(\omega-\omega'). 
    \label{eqn:mwmw}%
\end{equation}
where $\avg{\cdots}$ denotes the thermal ensemble average. The corresponding random fields are white:
\begin{equation}
    \Avg{b_i(t) b_j(t')} = \frac{2\alpha k_BT_F}{\gamma M}\delta_{ij}\delta(t-t'). 
    \label{eqn:hh}
\end{equation}
The auto-correlation for the interface-induced random field $\bb'$ has the form \Eq{eqn:hh} but with $\alpha\ra\alpha'$ and $T_{F}\ra T_{N}$.

We are interested in the DC (ensemble or time averaged) spin and energy currents across the FI$|$N interface. The transverse components average to zero, while the $z$-component in \Eq{eqn:llg0} has a bulk contribution driven by $B_{\rm app}$ and interfacial contribution $j_{s}^{z}$ driven by $(B_{\rm app},V_{s},\Delta T)$:
\begin{equation}
    \frac{\dM_z}{\gamma A} =
    -\frac{M}{\gamma A}\alpha_{0} \omega_{\rm app} \Avg{\mb\times(\mb\times\hzz)}_{z}+j_{s}^{z},
\end{equation}
where $\avg{\cdots}_{z}$ is the thermal average of the $z$-component,
\begin{align}
    j_s^z &= \frac{M}{\gamma A}
    \left\langle \alpha'\mb\times\dbm - \gamma\mb\times\bb' \right.  \nn
    &\qquad \left.  - \alpha'\smlb{\omega_s+\omega_{\rm app}}\mb\times(\mb\times\hzz)\right\rangle_{z}
\label{eqn:js}%
\end{align}
with $\dbm \approx -\gamma\mb\times\bB_{0}$. The associated energy current
equation
\begin{align}
    j_{Q} &= \frac{M}{\gamma A}
    \left\langle \alpha'\dbm\cdot\dbm - \gamma\dbm\cdot\bb' \right.  \nn
	& \qquad \left. - \alpha'\smlb{\omega_s+\omega_{\rm app}}\hzz\cdot(\mb\times\dbm) \right\rangle
\label{eqn:jQ}
\end{align}
follows from the interface contribution (terms proportional to $\alpha'$ and $\bb'$) of the energy change rate: $dE/dt = (d/dt) \avg{-\bB_0\cdot\bM}$. At zero temperature, $\mb = \hbz$ and $\bb' = \dbm = 0$, and both $j_s^z$ and $j_Q$ vanish, as expected. At finite temperatures, and in spite of $ \avg{\mb} \| \hbz$, $j_s^z$ and $j_Q$ are finite because $\mb$ and $\bb'$ are correlated. The relevant equal-time correlators $\avg{m_im_j},\avg{m_ib'_j},\avg{m_i\dm_j}$, and $\avg{\dm_ib'_j}$ can be derived from:
\begin{subequations}
\label{eqn:mmmh}
\begin{align}
    \avg{m_i(t)m_j(0)}  &= \frac{\gamma k_BT_F}{M} 
    \int\frac{d\omega}{2\pi}\frac{e^{-i\omega t}}{i\omega}\smlb{\chi - \chi^\dagger}_{ij},\\
    \avg{m_i(t)\gamma b'_j(0)} &= \frac{2\alpha'\gamma k_BT_N}{M}
    \int\frac{d\omega}{2\pi}e^{-i\omega t}\chi_{ij}.
\end{align}
\end{subequations}
Plugging these into \Eqs{eqn:js}{eqn:jQ}, we arrive at the linear response relation
\begin{equation}
    \smatrix{{\dM_z\ov\gamma A} \\ j_s^z \\ j_Q}
    = \frac{2\alpha'}{A} \frac{k_BT}{\omega_0} \smatrix{ 
	{\alpha\ov\alpha'} & 1 & \omega_0 \\ 
	1 & 1 & \omega_0 \\ 
	\omega_0 & \omega_0 & \omega_0^2 }
    \smatrix{ \omega_{\rm app} \\ \omega_s \\ {\Delta T\ov T}}.
\label{eqn:onsager_macro}
\end{equation}

\textit{Magnon model - } The macrospin model above holds only in the presence of strong applied magnetic fields or nanomagnets smaller than the exchange length. Otherwise the thermal spin wave excitations and magnetization texture $\mb(\br,t)$ ($\td{m}_{i}^{\bk,\omega}$ in Fourier space) may not be disregarded. The internal magnetic field should then be augmented by the exchange interaction $\bB_0\ra\bB_0+(D/\gamma\hbar)\nabla^{2}\mb$ where $D$ is the spin wave stiffness. The stochastic fields $\bb_0(\br,t)$ and $\bb'(\br,t)$ then depend on position and  $\bb'(\br,t)=\bar{\bb}'(y,z,t)\delta(x)$ acts at the interface. After linearizing and Fourier transforming \Eq{eqn:llg0} into frequency and momentum space, $\td{m}_i^{\bk,\omega} = \chi_{ij}^{\bk,\omega}\gamma\td{b}_j^{\bk,\omega}$, where the magnetic susceptibility $\chi^{\bk,\omega}$ takes the form \Eq{eqn:chi} with $\omega_0\ra\omega_0+(D/\hbar)k^{2}$. According to the FDT, the equilibrium magnetization fluctuations satisfy \cite{landau_statistical_1980}:
\begin{equation}
    {\avg{m_i^{\bk,\omega}m_j^{\bk',\omega'}}} = \frac{\gamma\hbar}{M_s}
    \frac{i\smlb{\chi^{\dag}-\chi}_{ij}}{e^{\hbar\omega/k_BT_F}-1} 
    \delta_{\omega\omega'}^{\bk\bk'},
    \label{eqn:mm}
\end{equation}
where $\delta_{\omega\omega'}^{\bk\bk'} \equiv \delta(\omega-\omega') \delta(\bk-\bk')$. The correlations for $\bb_0$ and $\bb'$ (or $\bar{\bb}'$) can be inferred from \Eq{eqn:mm}. The Planck distribution regulates the frequency integral over the continuum of magnon density of states. We note that the magnetic field dependence of the spin Seebeck effect at room temperature indicates a magnon cut-off lower than $k_BT_F/\hbar$. \cite{boona_magnon_2014,kikkawa_critical_2015,ritzmann_magnetic_2015}. 

The spin and energy currents across the FI$|$N interface are still given by \Eqs{eqn:js}{eqn:jQ} when substituting $\alpha'\ra\bar{\alpha}' = \alpha'd$, $\bb'\ra\bar{\bb}'$, and $M\ra M/d = M_sA$. Using \Eq{eqn:mm}, all equal-time-positon correlators in \Eqs{eqn:js}{eqn:jQ} can be infered from: 
\begin{subequations}
\label{eqn:mmmh2}
\begin{align}
    \avg{m_i(0,0)m_j(0,0)} &= n_FJ_{0}(x_0)\delta_{ij},\\
    \avg{m_i(0,0)\gamma \bar{b}'_j(0)}   &= -\bar{\alpha}'\frac{T_N}{T_F} \avg{\dm_i(0,0)m_j(0,0)},
\end{align}
\end{subequations}
where $n_F = \gamma\hbar/M_s\lambda^{3}$ is the total number of spins in the volume $\lambda^{3}=(4\pi D/  k_{B}T)^{3/2}$ and $\lambda$ is the de Broglie thermal wave length for magnons. $J_{l}(x_0) = \int_{0}^{\infty} 2\sqrt{x/\pi}(x_{0}+x)^{l}/(e^{x_{0}+x}-1)dx$ with $x_{0} =\hbar\omega_0/k_BT_F$. The expressions hold to leading order in $\alpha$. In the classical limit $x_0 \ll 1$, $J_l\ra Z_{l+3/2}$ with $Z_{n}$ the Zeta function. Using these correlators in \Eqs{eqn:js}{eqn:jQ} leads to the central result of this paper:
\begin{equation}
    \smatrix{{\dM_z\ov\gamma A} \\ j_s^z \\ j_Q} = \frac{2\bar{\alpha}'\hbar}{\lambda^3} \smatrix{
       	{\alpha\ov\alpha'}Z_{3\ov 2} & Z_{3\ov 2} & {3\ov 2\beta\hbar}Z_{5\ov 2} \\ 
	Z_{3\ov 2} & Z_{3\ov 2} & {3\ov 2\beta\hbar}Z_{5\ov 2} \\ 
	{3\ov 2\beta\hbar}Z_{5\ov 2} & {3\ov 2\beta\hbar}Z_{5\ov 2} & {15\ov 4\beta^2\hbar^2}Z_{7\ov 2} }
    \smatrix{\omega_{\rm app} \\ \omega_s \\ {\Delta T\ov T}}
    \label{eqn:onsager}
\end{equation}
with $\beta^{-1} = k_BT_F$. The response matrix is symmetric as required by Onsager reciprocity and invertible, \ie the forces are linear-independent as long as $\alpha_{0} \neq0$. 

\textit{Discussion - } The spin Seebeck effect represented by $L_{sQ}$:
\begin{equation}
    j_s^z 
    = 3Z_{5\ov 2}\frac{\bar{\alpha}'}{\lambda^3} k_B\Delta T 
    \simeq \frac{\Delta T}{0.1\,\mathrm{K}} \frac{0.14\mu\mathrm{J}}{\mathrm{m}^{2}}, 
    \label{eqn:sse}
\end{equation}
specifies the longitudinal spin current induced by the temperature difference $\Delta T$ \cite{xiao_theory_2010, adachi_linear-response_2011, hoffman_landau-lifshitz_2013}. The numerical estimates here and in the following are for the Pt$|$YIG system at $T=300$ K with parameters given in Table~\ref{tab:param}. The inverse of the spin Seebeck effect is the spin Peltier effect given by $L_{Qs}$:
\begin{equation}
    j_{Q} 
    = 3Z_{5\ov 2}\frac{\bar{\alpha}'}{\lambda^3} k_{B}T\frac{2e}{\hbar}V_{s} 
    \simeq\frac{V_{s}}{0.1\,\mathrm{\mu V}} \frac{1.3\times10^{5}\mathrm{J}}{\mathrm{m}^{2}\cdot\mathrm{s}}.
\label{eqn:spe}
\end{equation}
$V_{s}$ drives an energy current that cools/heats the magnons \cite{flipse_observation_2014}. 

The spin current driven by spin accumulations and external magnetic field
\begin{equation}
    j_s^z 
    = L_{sm}\omega_{\rm app}+L_{ss}\omega_{s}
    =2Z_{3\ov 2}\frac{\bar{\alpha}'}{\lambda^3}\hbar
    \smlb{ \gamma B_{\rm app} + {2e\ov\hbar}V_s} \label{eqn:sps}%
\end{equation}
vanishes with temperature since we disregarded quantum fluctuations and the magnon chemical potential in the ferromagnet \cite{bender_interfacial_2015}.{ $L_{sm}$ in \Eq{eqn:sps} quantifies the spin current induced by shifting the spin wave gap or ``squeezing'' the magnon distribution function. The spin conductance $L_{ss}$ in \Eq{eqn:sps} describes the spin current injection by a collinear spin accumulation $V_{s}$ as observed recently by Cornelissen \etal\cite{cornelissen_long_2015}. The magnon accumulation rate induced by this spin injection is described by $L_{ms}$. }

The energy current driven by the external field via $L_{Qm}$
\begin{equation}
    j_Q 
    = 3Z_{5\ov 2}\frac{\bar{\alpha}'}{\lambda^3}k_BT\gamma B_{\rm app} 
    \simeq\frac{B_{\rm app}}{1\,\mathrm{T}}\frac{7.4\times10^{7} \mathrm{J}}{\mathrm{m}^{2}\cdot\mathrm{s}}.
    \label{eqn:mpe}
\end{equation}
reflects what we call a {\it magnon Peltier effect}. It can be observed by a temperature change of the ferromagnet generated by the applied magnetic field, analogous to the spin Peltier effect caused by a spin accumulation \cite{flipse_observation_2014}. The reciprocal to the magnon Peltier effect is the magnetization dynamics induced by a temperature gradient via $L_{mQ}$
\begin{equation}
    \frac{\dM_z}{\gamma A}
    = 3Z_{5\ov 2}\frac{\bar{\alpha}'}{\lambda^3} k_B\Delta T,
\label{eqn:tf}%
\end{equation}
similar to applying a magnetic field. The magnitude of this {\it thermal effective magnetic field} can be estimated by setting $\dM_z=0$:
\begin{equation}
    B_{\Delta T} 
    = -\frac{\alpha'}{\alpha} \frac{3Z_{5/2}}{2Z_{3/2}} \frac{k_B\Delta T}{\gamma\hbar}
    = \frac{\bar{\alpha}'}{\bar{\alpha}'+\alpha_0d} \frac{\Delta T}{0.1\,\mathrm{K}}60\,\mathrm{mT},
\end{equation}
where the prefactor is of the order of unity as long as $d\ll\bar{\alpha}'/\alpha_0$ ($\sim 1~\mathrm{\mu}$m for YIG).
\begin{table}[t]
\centering
\begin{tabular}
[c]{c|l|c|l}\hline
Parameter & Value & Unit & Reference \hhline 
$M_{s}$ & $1.4\times10^{5}$ & A/m & \cite{dorsey_epitaxial_1993}\\
D & $8.5\times10^{-40}$ & J$\cdot$m$^{2}$ & \cite{cherepanov_saga_1993,srivastava_spin_1987} \\
$\alpha_{0}$ & $10^{-4}$ &  &  \\
$g_{r}$ & $10^{19}$ & 1/m$^{2}$ & \cite{weiler_experimental_2013} \\\hline
$\lambda$ & $1.6$ & nm & \\
$\bar{\alpha}^{\prime}$ & $0.1$ & nm & \\\hline
\end{tabular}
\caption{Parameters for YIG and YIG$|$Pt interface. }%
\label{tab:param}%
\end{table}

The conventional Seebeck coefficient refers to the open circuit thermopower voltage induced by a temperature difference. We may analogously define a spin Seebeck coefficient $S_{S}$ in terms of the spin thermopower, \ie the spin accumulation generated by a temperature bias $\Delta T$ in a spin-open circuit: $S_S=\smlb{V_s/\Delta T}_{j_s^z=0}$. This coefficient is not very relevant for metals that act as spin sinks such as Pt, but it characterizes the efficiency for weakly spin-dissipating metals such as Cu \cite{cahaya_spin_2014}. The thermally induced spin accumulation at zero spin current reads:
\begin{equation}
    V_s = -\frac{3Z_{5/2}}{4Z_{3/2}}\frac{k_B}{e}\Delta T \Longrightarrow
    S_S = -z\frac{k_B}{e},
\end{equation}
with $z=3Z_{5/2}/4Z_{3/2}\simeq0.385$. For single (or multiple) parabolic magnon bands $S_{S}\simeq33\,\mathrm{\mu}$V/K is a universal constant that does not depend on the spin wave stiffness, uniaxial anisotropy, and (in the considered regime) temperature.

The spin-magnonic interfacial thermal conductance $\kappa_m = (j_Q/\Delta T)_{j_s^z = 0} = (L_{QQ} - L_{sQ}L_{Qs}L_{ss}^{-1})/T$:
\begin{equation}
    \kappa_m 
    = z'\frac{\bar{\alpha}'}{\lambda^3} \frac{k_B^2T_F}{\hbar} \nn
    \simeq 0.76\times10^{8}\frac{\mathrm{W}}{\mathrm{m}^{2}\cdot\mathrm{K}}
\end{equation}
with $z'=(15Z_{7/2}-9Z_{5/2}^2/Z_{3/2})/2$, is of the same order of magnitude as the electron-phononic thermal conductance for Al$|$Al$_{2}$O$_{3}$ interface of $\sim 2\times10^{8}~\mathrm{W/}(\mathrm{m}^{2}\cdot\mathrm{K})$ \cite{cahill_nanoscale_2003}.

We can estimate the time scale of the transient response to a suddenly switched-on magnetic field. According to the first matrix element in \Eq{eqn:onsager} with $\omega_{\rm app} = \gamma B_{\rm app}$
\begin{equation}
    \avg{\dot{m}_{z}} = \frac{2\gamma\hbar}{M_s\lambda^3} Z_{3\ov 2}\alpha_0 \omega_{\rm app}
    = 2n_FZ_{3\ov 2}\alpha_0\omega_{\rm app}.
\end{equation}
On the other hand, the value of $m_{z}$ at thermal equilibrium depends on the spin wave gap $\hbar\omega_{0}$:
\begin{equation}
    \avg{m_{z}(\omega_{0})} = \sqrt{1-\avg{m_x^2}-\avg{m_y^2}} \simeq 1-n_FJ_0(x_0),
\end{equation}
that is shifted by $\omega_{\rm app}$, therefore
\begin{equation}
    \delta m_{z} = \frac{d\avg{m_{z}(\omega_{0})}}{d\omega_0}\omega_{\rm app}
    = n_F~\mbox{Li}_{\half}(x_0)\frac{\hbar\omega_{\rm app}}{k_BT},
    \label{eqn:dmz}
\end{equation}
where $\mbox{Li}$ is the PolyLog function. The magnon relaxation time to the
new equilibrium is therefore
\begin{equation}
    \tau \simeq \frac{\delta m_z}{\avg{\dm_z}}
    = \frac{\hbar}{\alpha k_BT} \frac{\mbox{Li}_{\half}(x_0)}{2Z_{3\ov 2}}
    \approx \frac{\hbar}{\alpha \sqrt{k_BT\hbar\omega_0}} \frac{\sqrt{\pi}}{2Z_{3\ov 2}},
    \label{eqn:tau}
\end{equation}
where the approximate expression is valid for $x_0 = \hbar\omega_0/k_BT\ll 1$. For YIG with $f = \omega_{0}/2\pi = 10$ GHz at room temperature, $\tau\simeq 2$ ns. Judging from the experimental magnetic field dependence of the spin Hall effect at room temperature, $k_BT$ should be replaced by a magnon spectrum cut-off of $~30$ K \cite{boona_magnon_2014,kikkawa_critical_2015,ritzmann_magnetic_2015}, leading to the estimate $\tau\sim 7$ ns. At even lower temperatures, $k_BT$ in \Eq{eqn:tau} should be replaced by $\hbar\omega_0$, leading to an upper estimate of $50$ ns.  

The non-equilibrium magnetization $M_s\delta m_z$ in \Eq{eqn:dmz} can be interpreted as a non-equilibrium magnon accumulation and the magnetic field as its driving force. This interpretation provides the link to theories of the electrically or thermally injected magnon Bose condensate in which the self-organized magnon chemical potential plays a crucial role \cite{bender_electronic_2012,duine_spintronics_2015}. The weak magnon-phonon interaction reported recently by Cornelissen \etal. \cite{cornelissen_long_2015} is encouraging that a long lived magnon chemical potential and condensate can be electrically generated. The pulsed magnetic field experiments suggested here do not require such a chemical potential and should yield useful insights into the magnon distribution function. 


In conclusion, we studied the magnetization dynamics coupled to spin and energy currents through FI$|$N interfaces at finite temperature as driven by collinear magnetic fields, spin accumulations, and/or a temperature bias. The response in this configuration vanishes with thermal fluctuations of the magnetization. We establish the Onsager reciprocal relations for these response functions. The elements in the Onsager matrix are identified as the spin Seebeck effect, the spin Peltier effect, and previously overlooked ones such as the magnon Peltier effect, effective thermal field, and magnon squeezing that have still to be observed experimentally. We identified a (nearly) universal spin Seebeck thermopower of $33~\mathrm{\mu}$V/K.

G.B. is grateful for the the hospitality of Rembert Duine at Utrecht University and his helpful explanations of the concept of the magnon chemical potential. J.X. thanks Yaroslav Tserkovnyak for the helpful discussion on the magnon relaxation time. Bart van Wees importantly helped us to understand the experiments by Cornelissen \etal \cite{cornelissen_long_2015}. This work was supported by the National Natural Science Foundation of China (11474065), National Basic Research Program of China (2014CB921600), the Foundation for Fundamental Research on Matter (FOM), DFG Priority Programme 1538 "Spin-Caloric Transport", JSPS\ Grant-in-Aid for Scientific Research (Nos. 25247056, 25220910, 26103006), and EU-FET Grant InSpin 612759.

\bibliographystyle{apsrev}

\end{document}